\documentclass[twocolumn,showpacs,floatfix,preprintnumbers,prl]{revtex4-1}
\usepackage{graphicx}
\usepackage{float}
\usepackage{dcolumn}
\usepackage{bm}
\usepackage{amssymb}

\begin{document}

\title{Interplay between electron-phonon couplings and disorder strength on the transport properties of organic semiconductors}

\author{C. A. Perroni and V. Cataudella}
\affiliation{CNR-SPIN and Dipartimento di Scienze Fisiche, Univ. di Napoli ``Federico II'', I-80126 Italy}

\begin{abstract}
The combined effect of bulk and interface electron-phonon couplings on the transport properties is investigated in a model for organic semiconductors gated with polarizable dielectrics. While the bulk electron-phonon interaction affects the behavior of mobility in the coherent regime below room temperature, the interface coupling is dominant for the activated high $T$ contribution of localized polarons. In order to improve the description of the transport properties, the presence of disorder is needed in addition to electron-phonon couplings. The effects of a weak disorder largely enhance the activation energies of mobility and induce the small polaron formation at lower values of electron-phonon couplings in the experimentally relevant window $150 K<T<300 K$.  The results are discussed in connection with experimental data of rubrene organic field-effect transistors.
\end{abstract}
\maketitle

In the last years, part of condensed matter research has focused on the realization of field-effect devices with correlated oxides and organic materials. \cite{RMP} This has opened up the possibility of injecting charge carriers without changing the chemical composition of the system, so that exploring phase diagrams of complex materials with small amount of disorder. On the other hand, the devices work with solid or liquid dielectric gates, therefore drawbacks arise due to different effects: for example, the lattice mismatch between bulk and gate, the interaction of bulk electrons with gate degrees of freedom, and the dielectric breakdown for strong electric fields. Finally, another important factor is the presence of traps at the interface between bulk and gate.

Single crystal organic field-effect transistors (OFETs) have represented a step forward in the field of plastic electronics. They are even more important since the intrinsic properties of organic materials can be explored. For example, single-crystal OFETs show a charge mobility at least one order of magnitude larger than that of thin films. \cite{hasegawa} Among them, the most promising and studied are those based on oligoacenes, like rubrene. \cite{morpurgo}

Scaling laws of the mobility as a function of the dielectric constant of solid \cite{stassen,nature} and liquid \cite{ono} gates have been discovered pointing out that the nearby dielectric has a strong influence. Actually, if the difference between the dielectric constant of the bulk and the gate is small, at temperatures close or higher than $100 K$, the mobility $\mu$ of these systems exhibits a power-law behavior ($\mu \sim T^{-\gamma}$, with $\gamma \simeq 2$). \cite{morpurgo} This conduction process cannot be simply ascribed to band transport,\cite{cheng,corop} and it arises from the interaction of charge carriers with bulk low frequency intermolecular modes. \cite{troisi_prl,vittoriocheck}  On the other hand, if the dielectric constant mismatch is high, an insulating behavior is found with much smaller values of mobility at room temperature. \cite{nature} It has been proposed that the injected charge carriers undergo a polaronic localization due to the interaction with the polarizable gate dielectrics. \cite{nature,fratini} However, accurate calculations have shown that the hopping barriers of localized polarons are too small with respect to the predicted values. \cite{bussac,bussac1} For example, the hopping barrier of rubrene transistors with a $Ta_2 O_5$ oxide gate is calculated to be $12.4$ meV, a value much lower than $55$ meV found in previous estimates. \cite{nature} Higher hopping barriers would require unrealistically large electron-phonon (el-ph) couplings.

Recently, the interplay between charge carrier coupling with intermolecular vibrations in the bulk and the long-range interaction induced at the interface with the dielectric gate has been investigated. \cite{giulio} These interactions stabilize a polaronic state, the bond polaron, at much lower strengths than expected from the polar interaction alone. However, this state still requires values of the bulk el-ph coupling larger than those typically estimated. Moreover, the transport properties with combined el-ph couplings have not yet been calculated and compared with experiments.

In this work, we investigate the combined effects of bulk and interface el-ph couplings on the transport properties in a regime of realistic values
of bulk el-ph interaction at finite temperatures. We show that the bulk coupling affects the behavior of mobility below room temperature enhancing the coherent contribution, but it is ineffective on the incoherent high $T$ contribution dominated by the interface coupling. Furthermore, we include the effect of a weak disorder due to bulk and interface traps. Actually, the interplay between long-range el-ph interactions and disorder effects is a largely unexplored issue. The disorder effects are able to enhance the hopping barriers of the activated mobility and to drive the small polaron formation at lower values of el-ph interactions. We point out that disorder is a key factor to get agreement with experimental data in rubrene OFETs.

{\it Model}
We study a one-dimensional Hamiltonian model with coupling to bulk and interface vibrational modes \cite{giulio}
\begin{equation}
H= H_{el}+H_{Bulk}^{(0)}+H_{el-Bulk}+H_{Inter}^{(0)}+H_{el-Inter}.
\label{h}
\end{equation}

In Eq. (\ref{h}), the electronic part $H_{el}$ is
\begin{equation}
H_{el} = -t \sum_{i}  \left( c_{i}^{\dagger} c_{i+1}+ h.c. \right) +  \sum_{i}  \epsilon_i n_{i} ,
\label{hel}
\end{equation}
where  $t$ is the bare electron hopping between the nearest neighbors on the chain with lattice parameter $a$, $c_{i}^{\dagger}$ and $c_{i}$ are the charge carrier creation and annihilation operators, respectively, relative to the site $R_i$, $\epsilon_i$ is a local energy whose fluctuations in the range $[-W,W]$ simulate disorder effects in the bulk and at the interface with gate, $ n_i= c_{i}^{\dagger} c_{i}$ is the density operator. The transfer hopping $t$ is estimated to be among $80 meV$  and $120 meV$. \cite{corop} Due to the presence of shallow traps, \cite{morpurgo} disorder is not overwhelming and it is distributed according to the flat probability function $Q \left( \{ \epsilon_i \}  \right) $.

In Eq. (\ref{h}), $H_{Bulk}^{(0)}$ is the Hamiltonian of free intermolecular modes
\begin{equation}
H_{Bulk}^{(0)}= \sum_{i} \frac{{p}^2_{i}}{2 m}+ \sum_{i} \frac{ k x_{i}^{2}}{2},
\label{hintra}
\end{equation}
where $x_{i}$ and ${p}_{i}$ are the oscillator displacement and momentum of the bulk mode at the site $i$, respectively, $m$ the oscillator mass, $k$ the elastic constant and $\omega_{Bulk}=\sqrt{k/m}$ the mode frequency. The intermolecular modes are characterized by small energies ($\hbar \omega_{Bulk} \simeq 5-10$ meV, with $\hbar$ Planck constant) in comparison with the transfer hopping. \cite{corop,troisi_prl}

In Eq. (\ref{h}),  $H_{el-Bulk}$ represents the term similar to the Su-Schrieffer-Heeger \cite{SSH} (SSH) interaction for the coupling to intermolecular modes
\begin{equation}
H_{el-Bulk}= \alpha \sum_{i} (x_{i+1}-x_i) \left( c_{i}^{\dagger}c_{i+1}+ h.c.  \right),
\label{hcoupling1}
\end{equation}
with $\alpha$ coupling constant. In the adiabatic regime for bulk modes ($\hbar \omega_{Bulk} \ll t$), the dimensionless quantity $\lambda_{Bulk}=\alpha^2/4 k t $ fully provides the strength of the electron coupling to intermolecular modes. The typical values of $\lambda_{Bulk}$ are in the intermediate coupling regime (in this work, we take $\lambda_{Bulk}=0.1$ suitable for rubrene). \cite{troisi_prl,vittoriocheck}

In Eq. (\ref{h}), $H_{Inter}^{(0)}$ is the Hamiltonian of free interface phonons
\begin{equation}
H_{Inter}^{(0)}= \hbar \omega_{Inter} \sum_{q} a_q^{\dagger} a_q,
\label{hintra}
\end{equation}
where $\omega_{Inter}$ is the frequency of optical modes, $a_{q}^{\dagger}$ and $a_{q}$ are creation and annihilation operators, respectively, relative to phonons with momentum $q$.

In Eq. (\ref{h}), $H_{el-Inter}$ is the Hamiltonian describing the electron coupling to interface vibrational modes
\begin{equation}
H_{el-Intra}= \sum_{i,q} M_q n_i e^{i q R_i} \left( a_q + a_{-q}^{\dagger}   \right),
\label{hcoupling}
\end{equation}
where $M_q$ is the interaction el-ph term
\begin{equation}
M_q= \frac{g \hbar \omega_{Inter} }{\sqrt{L}} \sum_{i} e^{i q R_i} \frac{R_0^2}{R_0^2+R_i^2},
\label{hcoupling}
\end{equation}
with $g$ dimensionless coupling constant, $L$ number of lattice sites, and $R_0$ cut-off length of the order of the lattice spacing $a$. In order to quantify this coupling, we use the dimensionless quantity $\lambda_{Inter}=\sum_q M_q^2/2 \hbar \omega_{Inter} t$. In this work, we take $R_0=0.5 a$ and $\hbar \omega_{Inter}=0.5 t$. \cite{bussac}

In the following, we will use units such that $a=1$, $\hbar=1$, $e=1$, and Boltzmann constant $k_B=1$.
We will analyze systems in the thermodynamic limit (up to $L=32$) measuring energies in units of
$t \simeq 100$ meV.

{\it Method}
The temperature range relevant for transport properties is such that $\omega_{Bulk} < T$, therefore the dynamics of intermolecular bulk modes can be accurately treated as classical. \cite{meSSH} The electron dynamics is influenced by the bulk oscillator deformations whose probability function $P \left( \{ x_i \}  \right) $ is self-consistently calculated through a Monte-Carlo approach. \cite{vittoriocheck} For one particle in a lattice of $L$ sites, the probability function is very close to that of free harmonic oscillators. \cite{perroni1} At a fixed configuration of bulk displacements $\{ x_i \}$, Eq.(\ref{hcoupling1}) shows that the electron hopping depend on the specific pair.

At a fixed configuration of displacements $\{ x_l \}$ and local energies $\{ \epsilon_m \}$, the resulting inhomogeneous model can be diagonalized using the following basis states
\begin{equation}
U_I \left( \{ x_l \} , \{ \epsilon_m \}  \right) |n_{q_1},.,n_{q_L} > |i>,
\label{basis}
\end{equation}
where the unitary transformation $U_I \left( \{ x_l \}, \{ \epsilon_m \} \right)$
\begin{eqnarray}
&& U_I \left( \{ x_l \} ,\{ \epsilon_m \}  \right)= \nonumber\\
&& \exp{ \left[ \frac{1}{\sqrt{L}} \sum_{j,q} f_q\left( \{ x_l \}, \{ \epsilon_m \} \right) n_j e^{i q R_j} \left( a_q - a_{-q}^{\dagger}   \right) \right] }
\end{eqnarray}
provides an el-ph treatment equivalent to a generalized Lang-Firsov approach. \cite{lang}  For long-range el-ph interactions, this approach is very accurate in the intermediate to strong coupling regime. \cite{alexandrov,perroni} Through the variational parameters
$f_q\left( \{ x_l \}, \{ \epsilon_m \} \right)$, interface phonons are displaced from their equilibrium position to a distance proportional to the el-ph interaction. The parameters $f_q\left( \{ x_l \}, \{ \epsilon_m \} \right)$ can be self-consistently calculated or, without losing accuracy, expressed as $f_q\left( \{ x_l \}, \{ \epsilon_m \} \right)=a \left( \{ x_l \}, \{ \epsilon_m \} \right) (M_q/ \omega_{Inter})/(b \left( \{ x_l \}, \{ \epsilon_m \} \right) \cos (q)+1)$, with $a \left( \{ x_l \}, \{ \epsilon_m \} \right)$ and $b \left( \{ x_l \}, \{ \epsilon_m \} \right)$ configuration dependent variational quantities. \cite{vitmanga}
In Eq. (\ref{basis}), $|n_{q_1},.,n_{q_L} >$ denotes the phononic basis in the occupation number representation, and $|i>=c_{i}^{\dagger}|0>$, with $|0>$ vacuum state, is the basis state for the electron.

After the diagonalization at fixed displacements $\{ x_l \}$ and energies $\{ \epsilon_m \}$, the solution is obtained by minimizing the free energy. A central quantity of this work is the mobility as a function of the temperature.
The mobility $\mu \left( \{ x_l \} , \{ \epsilon_m \} \right)$ is determined starting from the real part of the conductivity
$Re[\sigma(\omega)] \left( \{ x_l \} ,\{ \epsilon_m \} \right)$, taking the limit of zero frequency and dividing for the particle density $1/L$ of the system. In the linear regime, the real part of the conductivity is derived from the Kubo formula \cite{mahan}
\begin{eqnarray}
&& Re[\sigma(\omega)] \left( \{ x_l \}, \{ \epsilon_m \} \right)= \nonumber\\
&& \frac{ \left( 1-e^{-\beta \omega} \right) }{2 \omega} \frac{1}{L} \int_{-\infty}^{\infty} d t e^{i \omega t}
\langle j^{\dagger}(t) j(0) \rangle,
\label{kubo}
\end{eqnarray}
where $\beta=1/T$, $A(t)$ operator in Heisenberg representation, $j$ is the current operator
\begin{equation}
j=-i \sum_{j,\delta} \delta c^{\dagger}_{j+\delta} c_{j} h_{j,\delta} \left( \{ x_l \} \right),
\end{equation}
$ h_{j,\delta} \left( \{ x_l \} \right)= t-\alpha \delta (x_{j+\delta}-x_j) $ denotes the generalized hopping and $\delta=-1,1$ indicates the nearest neighbors.

The procedure for the calculation of the expectation value of an observable $O$ is the following: first, one makes the average
$ \left\langle O \left( \{ x_l \}, \{ \epsilon_m \} \right) \right\rangle$ over the eigenstates and eigenvectors of the corresponding inhomogeneous model, then, over the distributions $P \left( \{ x_i \} \right)$  and $Q \left( \{ \epsilon_j \} \right)$ making the integral
\begin{equation}
\left\langle \left\langle O \right\rangle \right\rangle = \int  \prod_{i} d x_{i} \prod_{j} d \epsilon_{j}   P \left( \{ x_i \} \right)
 Q \left( \{ \epsilon_j \} \right)  \left\langle O \left( \{ x_l \}, \{ \epsilon_m \} \right) \right\rangle
\label{distri}
\end{equation}
by means of the Monte-Carlo procedure. \cite{vittoriocheck}

\begin{figure}[htb]
\flushleft
\includegraphics[height=0.37\textwidth,angle=0]{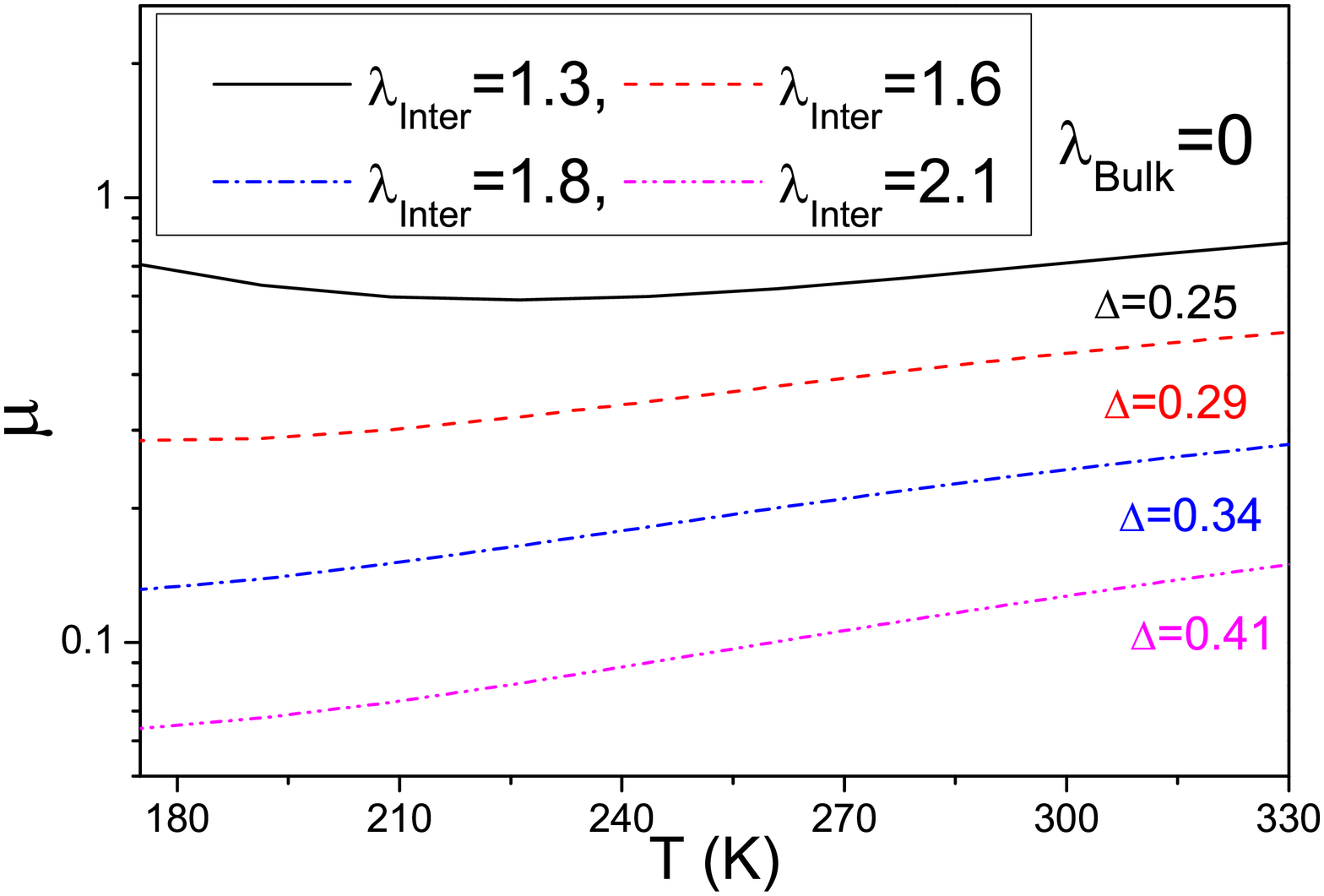}
\caption{Mobility (in units of  $\mu_0 = e a^2 / \hbar$) as a function of the temperature for different interface el-ph couplings $\lambda_{Inter}$ at $\lambda_{Bulk}=0$. The quantity $\Delta$ is the polaron activation energy.}
\label{spectral1}
\end{figure}

{\it Results}
First, we analyze the behavior of the mobility as a function of the temperature in the absence of bulk el-ph coupling and disorder. In Fig. \ref{spectral1}, we report the mobility for different values of $\lambda_{Inter}$ analyzing the intermediate to strong coupling regime. With increasing $\lambda_{Inter}$, the mobility decreases in the range of values smaller than unity. The mobility is expressed in natural units, that is in terms of $\mu_0 = e a^2 / \hbar
\simeq 7 cm^2 /(V \cdot s) $, taking $a=7 \AA$. \cite{corop} Therefore, the order of magnitude of the mobility is in agreement with data in rubrene OFETs for $1<\lambda_{Inter}<2$. \cite{nature} Below room temperature, the coherent metallic behavior is present only for $\lambda_{Inter} \leq 1.3$.
With increasing temperature, for $1<\lambda_{Inter}<2$, there is a hopping behavior with the activation energy $\Delta$: $\mu \propto \exp{(-\Delta/T)}$. Even for the unrealistically large coupling $\lambda_{Inter}=2.1$, the activation energy $\Delta=0.41 t \simeq 41$ meV is still smaller than that estimated in some rubrene OFETs. \cite{nature} This implies that further interactions have to be included in order to describe more accurately the mobility.

\begin{figure}[htb]
\centering
\includegraphics[height=0.36\textwidth,angle=0]{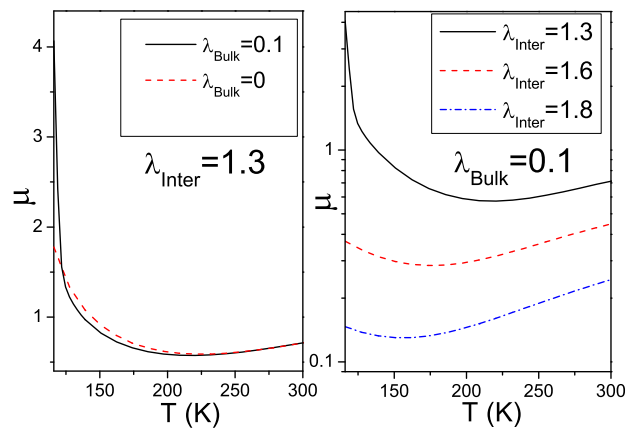}
\caption{Mobility (in units of  $\mu_0 = e a^2 / \hbar$) as a function of the temperature for different bulk el-ph couplings at $\lambda_{Inter}=1.3$ (Left panel) and for different interface el-ph couplings $\lambda_{Inter}$ at $\lambda_{Bulk}=0.1$ (Right panel).}
\label{spectral2}
\end{figure}

The next step is to analyze the role of bulk el-ph coupling. However, when one combines the bulk and interface el-ph couplings, the main effect is to improve the description of mobility in the coherent low temperature regime, leaving the high $T$ incoherent regime essentially unaltered for realistic bulk parameters. In the left panel of Fig. \ref{spectral2}, we report the mobility as a function of the temperature for $\lambda_{Inter}=1.3$ for bulk coupling $\lambda_{Bulk}=0$ and $\lambda_{Bulk}=0.1$ (appropriate to rubrene).
The physical situation corresponding to $\lambda_{Inter}=0$ has already been investigated in previous works. \cite{troisi_prl,vittoriocheck,nota}
As shown in Fig. \ref{spectral2},  $\mu$ shows a tiny coherent band-like behavior at low temperatures, but, with increasing $T$, it goes towards the activated behavior where the bulk coupling is not effective (in the right panel of Fig. \ref{spectral2} the polaron activation energies $\Delta$ are very close to those with $\lambda_{Bulk}=0$ shown in Fig. \ref{spectral1}). \cite{perroni1} Actually, the mobility interpolates between the behaviors with only one coupling. At low temperature, the diffusive contribution is ascribed to the modulation of the electron kinetic energy due to the bulk modes with SSH interactions.

In the right panel of Fig. \ref{spectral2}, the mobility as a function of the temperature is shown for different $\lambda_{Inter}$ at $\lambda_{Bulk}=0.1$.
A tiny metallic coherent behavior is present in the relevant temperature range for all the couplings. This coherent contribution is weakened with increasing
$\lambda_{Inter}$, but it does not disappear. This element is in contrast with experimental data which show a more or less marked insulating behavior from $150$ K to $300$ K. Therefore, the theoretical prediction of mobility is not accurate even if bulk and interface el-ph couplings are active.

\begin{figure}[htb]
\centering
\includegraphics[height=0.36\textwidth,angle=0]{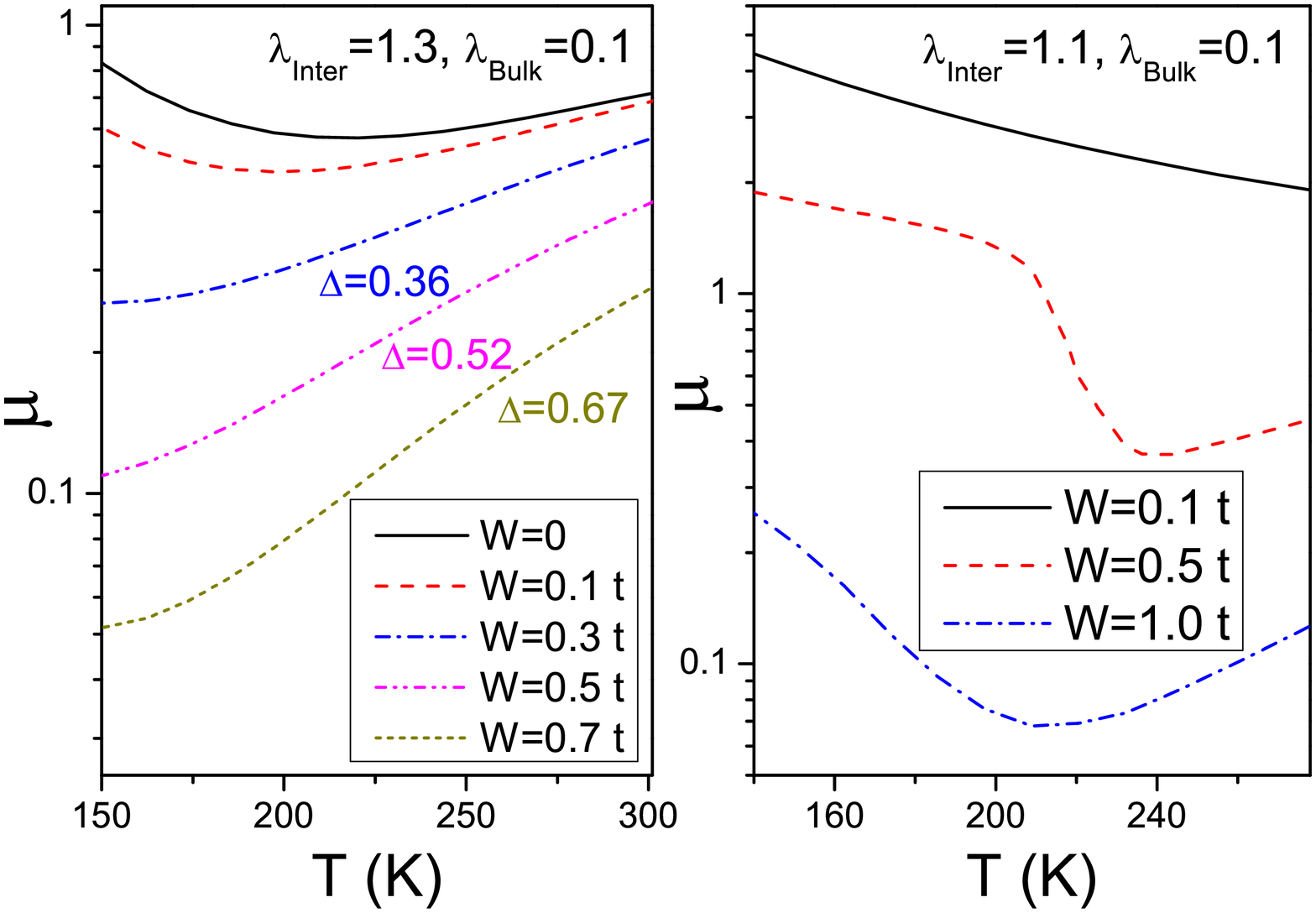}
\caption{Mobility $\mu$ (in units of  $\mu_0 = e a^2 / \hbar$)  as a function of the temperature for different disorder strengths $W$ and $\lambda_{Bulk}=0.1$ at $\lambda_{Inter}=1.3$ (Left panel) and $\lambda_{Inter}=1.1$ (Right panel). The quantity $\Delta$ is the polaron activation energy.}
\label{density1}
\end{figure}

In order to explain the experimental data, it is necessary to include also disorder effects. Indeed, there is evidence of traps in the bulk and at the interface with gates. \cite{morpurgo} In the left panel of Fig. \ref{density1}, we show the mobility as a function of the temperature with increasing the strength of disorder for $\lambda_{Inter}=1.3$ and $\lambda_{Bulk}=0.1$. There are two main results. The first one is related to the suppression of the coherent metallic behavior with increasing $W$. The second one is the strong enhancement of the activation energy $\Delta$ up to $0.67 t \simeq 67$ meV even for the small amount of disorder $W=0.7 t$. Furthermore, the decrease of the magnitude of the mobility is not so marked. Therefore, weak disorder effects are able to provide a very accurate description of the mobility resulting as key quantities for the interpretation of experimental data.

Another effect of disorder is to drive the small polaron formation at lower el-ph couplings. This phenomenon has already been studied in a single impurity model, \cite{alexandrov,hague} but, now, it is addressed in a more realistic situation where the disorder is distributed on all the lattice sites.  In the right panel of Fig. \ref{density1}, we show that the mobility as function of the temperature changes from metallic to insulating small polaron behavior at the lower interface el-ph coupling $\lambda_{Inter}=1.1$ and weak disorder $W=0.5 t$. At $W=1.0 t$, the coherent metallic term is strongly weakened, the transition temperature to localized polaron is decreased, and the activated regime is more stable.

In summary, the interplay between el-ph couplings and disorder effects influences the transport properties of organic semiconductors gated with polarizable dielectrics. The el-ph interactions for realistic values of model parameters are not sufficient to provide a satisfying description of mobility, so that we show that a weak disorder is able to fulfill the discrepancy between measured and calculated transport data. Disorder effects are able to reduce the coherent metallic behavior at low temperatures and to provide high temperature activation energies in agreement with data in rubrene organic field-effect transistors. The important point is that not only the interaction of bulk electrons with gate degrees of freedom, but also the presence of shallow traps in the bulk and at the interface is relevant for the properties of field-effect devices.


\begin{thebibliography}{99}
\bibitem{nota} For $\lambda_{Inter}=0$, $\mu$ scales as a power-law $\mu \propto T^{-\gamma}$, with $\gamma$ of the order of $2$.
For $T<150 K$, an upturn of the mobility as a function of the temperature can take place depending on the amount of disorder present in the bulk. This very simple effect due to traps will be discussed in a forthcoming longer paper.
\end{thebibliography}
\end{document}